\documentclass[preprint,prc,aps]{revtex4}
\everymath={\displaystyle}
\usepackage[latin1]{inputenc}
\usepackage{multirow}
\newcommand{\fr}{\frac}

\def\1{\mbox{l\hspace{-0.53em}1}}
\begin{document}
\title{The $[\bf{56},4^+]$ baryons in the $1/N_c$ expansion}

\author{N. Matagne\footnote{e-mail address: nmatagne@ulg.ac.be}}

\author{Fl. Stancu\footnote{ e-mail address: fstancu@ulg.ac.be
}}
\affiliation{University of Li\`ege, Institute of Physics B5, Sart Tilman,
B-4000 Li\`ege 1, Belgium}

\date{\today}

\begin{abstract}
\baselineskip=0.50cm
Using the $1/N_c$ expansion of QCD
we analyze the spectrum of positive parity resonances with strangeness
$S = 0, -1, -2$ and -3 in the 2-3 GeV mass region, supposed to belong 
to the $[\textbf{56},4^+]$ multiplet. The mass operator
is similar to that of  $[\textbf{56},2^+]$,
previously studied in the literature. The analysis of the latter is revisited. 
In the $[\textbf{56},4^+]$ multiplet we find that the spin-spin term brings the 
dominant contribution and that  the spin-orbit term is entirely negligible 
in the hyperfine interaction, in agreement with constituent quark model results. More data are strongly desirable, especially in the strange sector in order to fully exploit the power of this approach.
\end{abstract}

\maketitle
\section{Introduction}
The $1/N_c$ expansion of QCD \cite{tHo74,Wit79,GS84,bardacki} 
has been proved a useful approach 
to study baryon spectroscopy. It has been applied to the ground state
baryons \cite{DM93,DJM94,DJM95,CGO94,Jenk1,JL95,DDJM96} as well as to excited
states, in particular to the negative parity spin-flavor 
$[\textbf{70},1^-]$ multiplet 
($N = 1$ band)
\cite{Goi97,PY,CCGL,CGKM,CaCa98,SGS}, to the positive parity Roper resonance
belonging to $[\textbf{56'},0^+]$ ($N = 2$ band) \cite{CC00} and to the  
$[\textbf{56},2^+]$ multiplet ($N = 2$ band) \cite{GSS03}. In this approach 
the main features of the constituent quark model emerge naturally and in 
addition, new 
information is provided, as for example, on the spin-orbit problem.

In this study we explore its applicability to the $[\textbf{56},4^+]$ multiplet 
($N = 4$ band) for the first time.
The number of experimentally known resonances in the 2-3 GeV region
\cite{PDG04}, expected to belong to this 
multiplet is quite restricted. Among the five possible candidates
there are two four-star resonances, $N(2220) 9/2^+$ and 
$\Delta(2420) 11/2^+$, one three-star resonance 
$\Lambda(2350) 9/2^+$, one two-star resonance 
$\Delta(2300) 9/2^+$ and one one-star resonance
$\Delta(2390) 7/2^+$. This is an exploratory study 
which will allow us to make some predictions. 

In constituent quark models the $N = 4$ band has been studied so far either in 
a large harmonic oscillator basis \cite{CI} or in a 
variational basis \cite{SS}. We shall show that the present approach reinforces 
the conclusion that the spin-orbit contribution to the hyperfine interaction 
can safely be neglected in constituent quark model calculations.

The properties of low energy hadrons are interpreted to be a consequence of the
spontaneous breaking of chiral symmetry
\cite{GR}.
For highly excited hadrons, as the ones considered here, there are
phenomenological arguments 
to believe that the chiral symmetry is restored.
This would imply
a weakening (up to a cancellation) of the spin-orbit and tensor interactions  
\cite{Glozman}. Then   
the main contribution to the hyperfine interaction 
remains the spin-spin term.
\section{The wave functions}

The N = 4 band contains 17 multiples having symmetries (56), (70) or
(20)  and angular momenta ranging from 0 to 4 \cite{SS}. Among them, the  
$[\textbf{56},4^+]$ multiplet has a rather simple structure. It is symmetric both in
SU(6) and O(3), where O(3) is the group of spatial rotation. Together with
the color part which is always antisymmetric, it gives a totally antisymmetric
wave function. In our study of 
the $[\textbf{56},4^+]$ multiplet, we have to couple the symmetric orbital 
part $\psi^{\mathrm{Sym}}_{4m}$ with $\ell=4$ (see Table 2 of Ref. \cite{SS})
to a spin-flavor symmetric wave function. This gives
\begin{equation}
|4,S;d,Y,II_z; JJ_z\rangle = \sum_{m,\ S_z}\left(
                            \begin{array}{cc|c}
                                4 & S & J \\
                                m & S_z & J_z
                            \end{array}
       \right) \psi^{\mathrm{Sym}}_{4m} \times |SS_z;d,Y,II_z\rangle_{\mathrm{Sym}} ,
\label{states4+}
\end{equation}
where $S, S_z$ are the spin and its projection,
$d$  labels an SU(3) representation (here 8 and 10), $Y, I, I_z$  stand for
the hypercharge, isospin and its projection and $J, J_z$ for the total angular 
momentum and its projection. 
Expressing the states (\ref{states4+}) in the obvious notation
$^{2S+1}d_J$, they are as follows: two SU(3) octets
$^28_{\frac{7}{2}}$, $^28_{\frac{9}{2}}$ and four decuplets
$^410_{\frac{5}{2}}$, $^410_{\frac{7}{2}}$, $^410_{\frac{9}{2}}$,
$^410_{\frac{11}{2}}$.

In the following,  we need the explicit form of the wave
functions. They depend on  $J_z$ but the matrix elements of the
operators that we shall calculate in the next sections do not
depend on $J_z$ due to the Wigner-Eckart theorem. So, choosing
$J_z=\frac{1}{2}$, we have for 
the octet states
\begin{eqnarray}
|^28[\textbf{56},4^+]\frac{7}{2}^+\frac{1}{2}\rangle & = & \sqrt{\frac{5}{18}}\psi^S_{41}\left(\chi^{\rho}_-\phi^\rho+\chi^\lambda_-\phi^\lambda\right)-\sqrt{\frac{2}{9}}\psi^S_{40}\left(\chi^{\rho}_+\phi^\rho+\chi^\lambda_+\phi^\lambda\right), \label{wavefunc1}\\
|^28[\textbf{56},4^+]\frac{9}{2}^+\frac{1}{2}\rangle & = &\sqrt{\frac{2}{9}}\psi^S_{41}\left(\chi^{\rho}_-\phi^\rho+\chi^\lambda_-\phi^\lambda\right)+\sqrt{\frac{5}{18}}\psi^S_{40}\left(\chi^{\rho}_+\phi^\rho+\chi^\lambda_+\phi^\lambda\right),
\end{eqnarray}
and for the decuplet states
{\small 
\begin{eqnarray}
|^410[\textbf{56},4^+]\frac{5}{2}^+\frac{1}{2}\rangle  & = &  \left(\sqrt{\frac{5}{21}}\psi^S_{42}\chi_{\frac{3}{2}-\frac{3}{2}}-\sqrt{\frac{5}{14}}\psi^S_{41}\chi_{\frac{3}{2}-\frac{1}{2}}+\sqrt{\frac{2}{7}}\psi^S_{40}\chi_{\frac{3}{2}\frac{1}{2}}-\sqrt{\frac{5}{42}}\psi^S_{4-1}\chi_{\frac{3}{2}\frac{3}{2}}\right)\phi^S, \\
|^410[\textbf{56},4^+]\frac{7}{2}^+\frac{1}{2}\rangle & = & \left(\sqrt{\frac{3}{7}}\psi^S_{42}\chi_{\frac{3}{2}-\frac{3}{2}}-\sqrt{\frac{2}{63}}\psi^S_{41}\chi_{\frac{3}{2}-\frac{1}{2}}-\sqrt{\frac{10}{63}}\psi^S_{40}\chi_{\frac{3}{2}\frac{1}{2}}+\sqrt{\frac{8}{21}}\psi^S_{4-1}\chi_{\frac{3}{2}\frac{3}{2}}\right)\phi^S, \\
|^410[\textbf{56},4^+]\frac{9}{2}^+\frac{1}{2}\rangle & = & \left(\sqrt{\frac{3}{11}}\psi^S_{42}\chi_{\frac{3}{2}-\frac{3}{2}}+\sqrt{\frac{49}{198}}\psi^S_{41}\chi_{\frac{3}{2}-\frac{1}{2}}-\sqrt{\frac{10}{99}}\psi^S_{40}\chi_{\frac{3}{2}\frac{1}{2}}-\sqrt{\frac{25}{66}}\psi^S_{4-1}\chi_{\frac{3}{2}\frac{3}{2}}\right)\phi^S, \\
|^410[\textbf{56},4^+]\frac{11}{2}^+\frac{1}{2}\rangle & = & \left(\sqrt{\frac{2}{33}}\psi^S_{42}\chi_{\frac{3}{2}-\frac{3}{2}}+\sqrt{\frac{4}{11}}\psi^S_{41}\chi_{\frac{3}{2}-\frac{1}{2}}+\sqrt{\frac{5}{11}}\psi^S_{40}\chi_{\frac{3}{2}\frac{1}{2}}+\sqrt{\frac{4}{33}}\psi^S_{4-1}\chi_{\frac{3}{2}\frac{3}{2}}\right)\phi^S.
\label{wavefunc4}
\end{eqnarray}}
with
$\phi^\lambda$, $\phi^\rho$, $\phi^S$ and $\chi$  given in  Appendix \ref{firstapp}. 

\section{The mass operator}

The study of the $[\textbf{56},4^+]$ multiplet is similar to that of $[\textbf{56},2^+]$
as analyzed in Ref. \cite{GSS03}, where the mass
spectrum is studied in the $1/N_c$ expansion up to and including 
$\mathcal{O}(1/N_c)$ effects.
The mass operator must be rotationally invariant, parity and time reversal
even. The isospin breaking is neglected.  
The SU(3) symmetry breaking is implemented 
to $\mathcal{O}(\varepsilon)$,
where $\varepsilon \sim 0.3$ gives a measure of this breaking.
As the $[\textbf{56},4^+]$ baryons are 
described by a symmetric representation of SU(6), 
it is not necessary to distinguish between
excited and core quarks for the construction of a basis of mass operators,  
as explained in Ref. \cite{GSS03}. Then 
the mass operator of the $[{\bf 56},4^+]$ multiplet
has the following structure 
\begin{equation}
\label{MASS}
M = \sum_i c_i O_i + \sum_i b_i \bar{B}_i
\end{equation}
given in terms of the linearly independent operators $O_i$ and $\bar{B}_i$,
similar to that of the $[{\bf 56},2^+]$ multiplet. Here
$O_i$ ($i = 1,2,3$) are rotational invariants and SU(3)-flavor singlets
\cite{Goi97}, $\bar{B}_1$ is the strangeness quark number operator
with negative sign, 
and the operators $\bar{B}_i$ ($i = 2, 3$)
are also rotational invariants but contain the SU(6) flavor-spin generators 
$G_{i8}$ as well. The operators
$\bar{B}_i$ ($i = 1, 2, 3$) provide SU(3) breaking and are defined to
have vanishing matrix elements for nonstrange baryons.
The relation (\ref{MASS}) contains the effective 
coefficients $c_i$ and $b_i$ as parameters. They represent reduced matrix 
elements that encode the QCD dynamics.
The above operators and the values of the corresponding coefficients 
obtained from fitting the experimentally known masses (see next section) 
are given in Table \ref{operators}.

We recall that a generic $n$-body operator has the structure
\begin{equation}
O^{(n)}=\frac{1}{N_c^{n-1}}O_\ell O_{SF}
\end{equation}
where the factors $O_\ell$ and $O_{SF}$  can be expressed in terms of products 
of generators $\ell_i$ ($i = 1, 2, 3$) of the group O(3), and of the 
spin-flavor group SU(6) $S_i$, $T_a$ and 
$G_{ia}$ ($i=1, 2, 3$; $a=1, \ldots, 8$). Because an $n$-body operator requires
 that at least $(n-1)$ gluons be exchanged between $n$ quarks, an overall 
factor of $1/N_c^{n-1}$ appears. Matrix elements of some operators can carry 
a nontrivial $N_c$ dependence due to coherence effects \cite{DM93}: for example,
$G_{ia}$ $(a=1,2,3)$ and $T_8$ have matrix elements of $\mathcal{O}(N_c)$.
This explains the dependence on $N_c$ of the operators listed in
Table \ref{operators}.

The matrix elements of $O_1$, $O_2$ and $O_3$ are trivial 
to calculate. They are given in Table \ref{singlets} for the octet and
the decuplet states belonging to the $[{\bf 56},4^+]$ multiplet. The  
$\bar{B}_1$ matrix elements are also trivial.  To calculate
the $\bar{B}_2$ matrix elements 
we use the expression 
\begin{equation}\label{Gi8}
G_{i8} = G^{i8} = \frac{1}{2 \sqrt{3}}(S^i - 3 S^i_s), 
\end{equation}
where $S^i$ and $S^i_s$ are the components of the total spin 
and of the total strange-quark spin respectively \cite{JL95}.
We rewrite this expression as 
\begin{equation}\label{B2}
\bar{B}_2 = - \frac{\sqrt{3}}{2 N_c} \vec{l}\cdot \vec{S}_s
\end{equation}
with the decomposition 
\begin{equation}\label{decomposition}
\vec{l}\cdot \vec{S}_s = l_0S_{s0}+\frac{1}{2}\left( l_+S_{s-}+l_-S_{s+}\right),
\end{equation} 
which we apply on the wave functions 
(\ref{wavefunc1})--(\ref{wavefunc4}) in order to obtain the diagonal 
and off-diagonal matrix elements.  For $\bar{B}_3$, one can use the following 
relation \cite{CC00}
\begin{equation}\label{B3}
S_i G_{i8} = \fr{1}{4 \sqrt{3}} [3I(I+1) -S(S+1) - \fr{3}{4} N_s(N_s+2)]
\end{equation}
in agreement with \cite{JL95}. Here $I$ is the isospin, $S$ is the total
spin and  $N_s$ the number of strange quarks. Both the diagonal and
off-diagonal matrix elements of  $\bar{B}_i$  are exhibited in 
Table \ref{octets}. Note that only  $\bar{B}_2$ has non-vanishing off-diagonal
matrix elements. Their role is very important in the state mixing, as discussed
in the next section. We found that the diagonal matrix elements of $O_2$, $O_3$,
$\bar{B}_2$ and $\bar{B}_3$ of strange baryons satisfy the following relation
\begin{equation}\label{dependence}
\frac{\bar{B}_2}{\bar{B}_3} =\frac{O_2}{O_3},
\end{equation}
for any state, irrespective of the value of $J$ in both the octet and the
decuplet. This can be used as a check of the analytic expressions in Table 
\ref{octets}. Such a relation also holds for the multiplet $[\textbf{56},2^+]$
studied in Ref. \cite{GSS03} and might possibly be a feature of all
$[\textbf{56},\ell +]$ multiplets. In spite of the relation (\ref{dependence}) 
which holds for the diagonal matrix elements, the operators $O_i$ and $\bar{B}_i$ 
are linearly independent, as it can be easily proved. As a proof, 
the off-diagonal matrix elements of $\bar{B}_2$ are entirely different
from those of $\bar{B}_3$.

\section{State mixing}

As mentioned above, only the operator $\bar{B}_2$ has non-vanishing off-diagonal 
matrix elements, so $\bar{B}_2$ is the only one which 
induces mixing between the octet and decuplet states of $[\textbf{56},4^+]$
with the same quantum numbers, as a consequence of the SU(3)-flavor breaking.
Thus this mixing affects the octet and the decuplet $\Sigma$ and 
$\Xi$ states. As there are four off-diagonal matrix elements
(Table \ref{octets}), there are also 
four mixing angles, namely,  $\theta_J^{\Sigma}$ and $\theta_J^{\Xi}$, each with
$J$ =7/2 and 9/2.  In terms of these mixing angles, the physical $
\Sigma_J$ and $\Sigma_J'$ states are defined by the following basis states
\begin{eqnarray}
|\Sigma_J \rangle  & = & |\Sigma_J^{(8)}\rangle \cos\theta_J^{\Sigma} 
+ |\Sigma_J^{(10)}\rangle \sin\theta_J^{\Sigma}, \\
|\Sigma_J'\rangle & = & -|\Sigma_J^{(8)}\rangle \sin\theta_J^{\Sigma}
+|\Sigma_J^{(10)} \rangle \cos\theta_J^{\Sigma},
\end{eqnarray}
and similar relations hold for $\Xi$. 
The masses of the physical states become
\begin{eqnarray}
M(\Sigma_J) & = & M(\Sigma_J^{(8)}) 
+ b_2 \langle \Sigma_J^{(8)}|\bar{B}_2|\Sigma_J^{(10)} \rangle \tan \theta^{\Sigma}_J, \label{masssigmaj} \\
M(\Sigma_J') & = & M(\Sigma_J^{(10)}) 
- b_2 \langle \Sigma_J^{(8)}|\bar{B}_2|\Sigma_J^{(10)} \rangle \tan \theta^{\Sigma}_J \label{masssigma'j},
\end{eqnarray}
where $M(\Sigma_J^{(8)})$ and $M(\Sigma_J^{(10)})$ are the diagonal matrix 
of the mass operator (\ref{MASS}), 
here equal to $c_1O_1+c_2O_2+c_3O_3+b_1\bar{B}_1$, 
for $\Sigma$ states and  similarly for $\Xi$ states 
(see Table \ref{multiplet}). 
If replaced in the mass operator (\ref{MASS}), the relations (\ref{masssigmaj}) 
and  (\ref{masssigma'j}) and their counterparts for $\Xi$, introduce 
four new parameters which should be included in the fit.
Actually the procedure of Ref. \cite{GSS03} was simplified to fit the coefficients 
$c_i$ and $b_i$ directly to the physical masses
and then to calculate the mixing angle from
\begin{equation}\label{angle}
\theta_J = 
\frac{1}{2}\arcsin\left( 2~ \frac{b_2\langle \Sigma_J^{(8)}|\bar{B}_2|\Sigma_J^{(10)} \rangle}
{ M(\Sigma_J) - M(\Sigma'_J)}\right).
\label{mixingangles}
\end{equation} 
for $\Sigma_J$ states and analogously for $\Xi$ states.  

As we shall see below, due to the scarcity of
data in the 2-3 GeV mass region, even such a simplified procedure
is not possible at present in the $[\textbf{56},4^+]$ multiplet. 
\section{Mass relations}

In the isospin symmetric limit, there are twenty four independent masses, 
as presented in the first column of Table \ref{multiplet}. Our operator basis 
contains six operators, so there are eighteen mass relations that hold 
irrespective of the values of the coefficients $c_i$ and $b_i$. These relations 
can be easily calculated from the definition (\ref{MASS}) and are presented in 
Table  \ref{masrel}. 

One can identify the Gell-Mann--Okubo (GMO) mass formula for each octet (two such relations) 
and the equal spacing relations (EQS) for each decuplet (eight such relations). 
There are eight relations left, that involve states belonging to different SU(3)
multiplets as well as to different values of $J$. 
Presently one cannot test the accuracy of these relations due to lack of data.
But they may be used in making some predictions. The theoretical masses
satisfy the Gell-Mann--Okubo mass formula for octets and the Equal Spacing
Rule for decuplets, providing another useful test of the $1/N_c$ expansion.

\section{The $[\textbf{56},2^+]$ revisited}

We have checked the analytic work of Ref. \cite{GSS03} and refitted the
masses of the octets and decuplets with a slightly different input.  
This is related to a different assignement of the $\Delta_{5/2}$ resonances. 

The analysis performed in Ref. \cite{GSS03} is based on the standard 
identification of resonances due to the pioneering work of Isgur and Karl
\cite{IK}. In that work the spectrum of positive parity resonances is
calculated from a Hamiltonian containing a harmonic oscillator confinement
and a hyperfine interaction of one-gluon exchange type. The mixing angles
in the $\Delta_{5/2^+}$ sector turn out to be

\begin{center}
\renewcommand{\arraystretch}{1.25}
\begin{tabular}{ccc}
\hline \hline
State & Mass (MeV) &  Mixing angles\\
\hline
$^4\Delta[\textbf{56},2^+]\frac{5}{2}^+$ & $1940$ & \multirow{2}{2.65cm}{$\left[ \begin{array}{cc} 0.94  & 0.38 \\  -0.38 & 0.94 \end{array}\right]$} \\
$^4\Delta[\textbf{70},2^+]\frac{5}{2}^+$ & $1975$ & \\
\hline \hline
\end{tabular}
\end{center}
which shows that the lowest resonance at 1940 MeV is dominantly  
a $[\textbf{56},2^+]$ state. As a consequence, the lowest observed 
$F_{35}\ \Delta(1905)$ resonance is interpreted as a member of the $[\textbf{56},2^+]$ multiplet.

In a more realistic description, based on a linear confinement
\cite{SS86}, the structure of  the $\Delta_{5/2^+}$ sector appeared
to be different. The result was 

\begin{center}
\renewcommand{\arraystretch}{1.25}
\begin{tabular}{ccc}
\hline \hline
State & Mass (MeV) & Mixing angles \\
\hline
$^4\Delta[\textbf{56},2^+]\frac{5}{2}^+$ & 1962 & \multirow{2}{3.15cm}{$\left[ \begin{array}{cc} 0.408 & 0.913 \\ 0.913 & -0.408 \end{array}\right]$} \\
$^4\Delta[\textbf{70},2^+]\frac{5}{2}^+$ & $1985$ & \\
\hline \hline
\end{tabular}
\end{center}
which means that in this case the higher resonance of mass 1985 MeV is
dominantly $[\textbf{56},2^+]$. Accordingly, here we interpret the higher experimentally
observed resonance $F_{35}~ \Delta(2000)$ as belonging to the $[\textbf{56},2^+]$ multiplet
instead of the lower one. Thus we take as experimental input the mass
1976 $\pm$ 237 MeV, determined from the full listings of the PDG \cite{PDG04} 
in the same manner as for the one-
and two-star resonances of the $[\textbf{56},4^+]$
multiplet (see below). The $\chi^2_{\mathrm{dof}}$ obtained is 0.58, as
compared to  $\chi^2_{\mathrm{dof}}$ = 0.7 of Ref. \cite{GSS03}. 
Although $\Delta(2000)$ is a two-star resonance only, the incentive of
making the above choice was that the calculated pion decay 
widths of the $\Delta_{5/2^+}$ sector were better reproduced \cite{SS95} 
with the mixing angles of the
model \cite{SS86} than with those of the standard model of \cite{IK}.  
It is well known that decay widths are useful to test mixing angles. 
Moreover, it would be more natural that the resonances $\Delta_{1/2}$
and $\Delta_{5/2}$ would have different masses, contrary to the 
assumption of Ref. \cite{GSS03} where these masses are identical.

\section{Fit and discussion}

The fit of the  masses derived from  Eq. (\ref{MASS})
and the available empirical values used in the fit,
together with the corresponding resonance status in the Particle Data Group
\cite{PDG04} are listed in Table \ref{multiplet}.
The values of the coefficients $c_i$ and $b_1$ obtained from 
the fit are presented in Table \ref{operators}, as already mentioned.
For the four and three-star resonances we used the empirical masses given in 
the summary  table. For the others, namely the one-star resonance
$\Delta(2390)$ and the two-star resonance $\Delta(2300)$
we adopted the following procedure. We considered as "experimental" mass
the average of all masses quoted in the full listings. The experimental
error to the mass was defined as the quadrature of two uncorrelated errors,
one being the average error obtained from the same references in the 
full listings  and the other was the difference 
between the average mass relative to the farthest off observed mass. The 
masses and errors thus obtained are indicated in the before last column
of Table \ref{multiplet}.
   
Due to the lack of experimental data in the strange sector it was 
not possible to include all the operators $\bar{B}_i$ in the fit in order 
to obtain 
some reliable predictions.  As the breaking of SU(3) is dominated by 
$\bar{B}_1$ we included only this operator in  Eq. (\ref{MASS})
and neglected the contribution of the operators $\bar{B}_2$ and $\bar{B}_3$.
At a later stage, when more data will hopefully be available, all analytical 
work performed here could be used to improve the fit. That is why
Table \ref{operators} contains results for  $c_i$ ($i$ = 1, 2 and 3) 
and $b_1$ only. The $\chi^2_{\mathrm{dof}}$ of the fit is 0.26, where 
the number of degrees of freedom (dof) is equal to one (five data and four 
coefficients).

The first column of Table  \ref{multiplet}
 contains the 56 states (each state having a $2 I + 1$ multiplicity
from assuming an exact SU(2)-isospin symmetry) \footnote{%
Note that the notation $\Sigma_J$, $\Sigma'_J$ is consistent with 
the relations (\ref{masssigmaj}),~(\ref{masssigma'j}) inasmuch as 
the contribution of $\bar{B}_2$ is neglected (same remark 
for $\Xi_J$, $\Xi'_J$ and corresponding relations).}. 
The columns two to five show the 
partial contribution of each operator included in the fit, multiplied by
the corresponding coefficient $c_i$ or $b_1$.
The column six gives the total mass according to Eq. (\ref{MASS}). 
The errors shown in the predictions result from the errors on the 
coefficients $c_i$ and $b_1$ given in Table \ref{operators}.
As there are only five experimental data available, nineteen of these 
masses are predictions. The breaking of SU(3)-flavor due to the operator
$\bar B_1$ is 110 MeV as compared to 200 MeV produced in the 
$[\textbf{56},2^+]$ multiplet.

The main question is, of course, how reliable is this fit. The answer
can be summarized as follows:
\begin{itemize}
\item The main part of the mass is provided by the spin-flavor singlet operator
$O_1$, which is $\mathcal{O}(N_c)$. 

\item The spin-orbit contribution given by $c_2O_2$ is small. This fact 
reinforces the practice used in constituent quark models where the spin-orbit
contribution is usually neglected. Our result is consistent with the expectation
that the spin-orbit term vanishes at large excitation energies \cite{Glozman}.
  
\item The breaking of the $SU(6)$ symmetry keeping the flavor symmetry exact
is mainly due to the spin-spin operator $O_3$. This hyperfine interaction 
produces a splitting between octet and decuplet states of approximately 130 MeV 
which is smaller than that obtained in the $[\textbf{56},2^+]$ 
case \cite{GSS03}, which gives 240 MeV.

\item The contribution of $\bar{B}_1$
per unit of strangeness, 110 MeV, is also smaller here than in the 
$[\textbf{56},2^+]$ multiplet \cite{GSS03}, where it takes a value of about 
200 MeV. That may be quite natural, as one expects a shrinking of the spectrum
with the excitation energy.

\item As it was not possible to include the contribution of $\bar{B}_2$ and 
$\bar{B}_3$ in our fit, a degeneracy appears between $\Lambda$ and $\Sigma$. 

\end{itemize}

\section{Conclusions}

We have studied the spectrum of highly excited resonances in the 2-3 GeV
mass region by describing them as belonging to the $[\textbf{56},4^+]$ multiplet.
This is the first study of such excited states based on the $1/N_c$ 
expansion of QCD. A better description should include multiplet 
mixing, following the lines developed, for example, in Ref. \cite{Go04}. 
 
We support previous assertions that better experimental values 
for highly excited non-strange baryons as well as more data 
for the $\Sigma^*$ and $\Xi^*$ baryons are needed in order to understand
the role of the operator $\bar{B}_2$  within a multiplet
and for the octet-decuplet mixing. With better data the analytic work 
performed here will help 
to make reliable predictions in the large $N_c$ limit formalism.

\vspace{2cm}

{\bf Acknowledgments}.

We are grateful to  J.~L. Goity, C. Schat and N.~N. Scoccola for illuminating 
discussions during the Large $N_c$ QCD workshop at ECT* Trento, Italy 2004
and to P. Stassart
for useful advise in performing the fit. 
The work of one of us (N. M.) was supported by the Institut Interuniversitaire 
des Sciences Nucl\'eaires (Belgium).

\par

\vspace{1cm}
\appendix
\section{Flavor-spin states}
\label{firstapp}

Here we reproduce the definitions of the flavor and spin states used in 
the calculations of the matrix elements of ${\bar B}_2$ using the
standard notations \cite{stancu}.
In the $[\textbf{56},4^+]$ multiplet the spin-flavor part of the baryon 
wave function is symmetric (see relation (\ref{states4+})).
Consider first the octet. One can write
\begin{equation}
|SS_z; 8, Y, II_z\rangle_{\mathrm{Sym}}= \frac{1}{\sqrt{2}}\left( \chi^\rho \phi^\rho +\chi^\lambda \phi^\lambda\right)
\end{equation}
with the flavor states
$\phi^\rho$ and $\phi^\lambda$ defined in Table \ref{msymflstates}.
The states
$\chi^\rho$ and $\chi^\lambda$ represent of spin $S=1/2$ and permutation 
symmetry $\rho$ and $\lambda$ respectively, can be obtained from 
$\phi^\rho_{p(n)}$ and $\phi^\lambda_{p(n)}$ by making the replacement
$$u\rightarrow \uparrow, \ d\rightarrow \downarrow$$
where $\uparrow$ and $\downarrow$ are spin 1/2 single-particle states of projection $S_z=+1/2$ and $S_z=-1/2$, respectively. The three particle states of mixed symmetry and $S=1/2$, $S_z=+1/2$ are
\begin{equation}
\chi^\lambda_+=-\frac{1}{\sqrt{6}}\left(\uparrow\downarrow\uparrow+\downarrow\uparrow\uparrow-2\uparrow\uparrow\downarrow\right)
\end{equation}
\begin{equation}
\chi^\rho_+=\frac{1}{\sqrt{2}}\left( \uparrow\downarrow\uparrow-\downarrow\uparrow\uparrow\right)
\end{equation}
and those having $S=1/2$, $S_z=-1/2$ are
\begin{equation}
\chi^\lambda_-=\frac{1}{\sqrt{6}}\left(\uparrow\downarrow\downarrow+\downarrow\uparrow\downarrow-2\downarrow\downarrow\uparrow\right)
\end{equation}
\begin{equation}
\chi^\rho_-=\frac{1}{\sqrt{2}}\left(\uparrow\downarrow\downarrow-\downarrow\uparrow\downarrow\right).
\end{equation}
For the decuplet, we have
\begin{equation}
|SS_z; 10, Y, II_z\rangle_{\mathrm{Sym}}=\chi\phi^S
\end{equation}
with $\phi^S$ defined in Table \ref{symflstates}. $\chi$ must be an $S=3/2$ state. With the previous notation, the $\chi_{\frac{3}{2}m}$ states take the form
\begin{equation}
\chi_{\frac{3}{2}\frac{3}{2}}=\uparrow\uparrow\uparrow
\end{equation}
\begin{equation}
\chi_{\frac{3}{2}\frac{1}{2}}=\frac{1}{\sqrt{3}}\left( \uparrow\uparrow\downarrow+\uparrow\downarrow\uparrow+\downarrow\uparrow\uparrow\right)
\end{equation}
\begin{equation}
\chi_{\frac{3}{2}-\frac{1}{2}}=\frac{1}{\sqrt{3}}\left(\uparrow\downarrow\downarrow+\downarrow\uparrow\downarrow+\downarrow\downarrow\uparrow\right)
\end{equation}
\begin{equation}
\chi_{\frac{3}{2}-\frac{3}{2}}=\downarrow\downarrow\downarrow
\end{equation}



\pagebreak

\begin{table}[tbp]
\renewcommand{\arraystretch}{1.25}
\begin{tabular}{llrrl}
\hline
\hline
Operator & \multicolumn{4}{c}{Fitted coef. (MeV)}\\
\hline
\hline
$O_1 = N_c \ \1 $                                               & \ \ \ $c_1 =  $  & 736 & $\pm$ & 30      $\ $ \\
$O_2 =\frac{1}{N_c} l_i  S_i$                                  & \ \ \ $c_2 =  $  &  4 & $\pm$ & 40   $\ $ \\
$O_3 = \frac{1}{N_c}S_i S_i$                                    & \ \ \ $c_3 =  $  &  135 & $\pm$ & 90   $\ $ \\
\hline
$\bar B_1 = -{\cal S} $                                         & \ \ \ $ b_1 = $  & 110 & $\pm$ & 67   $\ $ \\
$\bar B_2 = \frac{1}{N_c} l_i G_{i8}-\frac{1}{2 \sqrt{3}} O_2$  & \ \ \  \\

$\bar B_3 = \frac{1}{N_c} S_i G_{i8}-\frac{1}{2 \sqrt{3}} O_3$  & \ \ \ &  & &  \\

\hline \hline
\end{tabular}
\caption{Operators of Eq. (\ref{MASS}) and coefficients 
resulting from the fit with $\chi^2_{\rm dof}  \simeq 0.26$.  }
\label{operators}
\vspace{4cm}
\end{table}


\begin{table}
\[
\renewcommand{\arraystretch}{1.5}
\begin{array}{crrr}
\hline
\hline
           & \ \ \ \ \ \ \ \  O_1  & \ \ \ \ \ \ \ O_2  & \ \ \ \ \ \ \ O_3  \\
\hline
^28_{7/2}  & N_c  & - \fr{5}{2 N_c} & \fr{3}{4 N_c}  \\
^28_{9/2}  & N_c  &   \fr{2}{  N_c} & \fr{3}{4 N_c}  \\
^410_{5/2} & N_c  & - \fr{15}{2 N_c} & \fr{15}{4 N_c} \\
^410_{7/2} & N_c  & - \fr{4}{  N_c} & \fr{15}{4 N_c} \\
^410_{9/2} & N_c  &   \fr{1}{2 N_c} & \fr{15}{4 N_c} \\
^410_{11/2} & N_c  &   \fr{6}{  N_c} & \fr{15}{4 N_c} \\[0.5ex]
\hline
\hline
\end{array}
\]
\caption{Matrix elements of  SU(3) singlet operators.}
\label{singlets}
\end{table}

\clearpage


\begin{table}
\[
\renewcommand{\arraystretch}{1.5}
\begin{array}{cccc}
\hline
\hline
  &  \hspace{ .6 cm}  {\bar B}_1 \hspace{ .6 cm}  &
     \hspace{ .6 cm}  {\bar B}_2 \hspace{ .6 cm}  &
     \hspace{ .6 cm}  {\bar B}_3 \hspace{ .6 cm}   \\
\hline
N_{J}       & 0 &                      0  &                           0  \\
\Lambda_{J} & 1 &   \fr{  \sqrt{3}\ a_J}{2 N_c} &   - \fr{3 \sqrt{3}}{8 N_c}   \\
\Sigma_{J}  & 1 & - \fr{  \sqrt{3}\ a_J}{6 N_c} &     \fr{  \sqrt{3}}{8 N_c}   \\
\Xi_{J}     & 2 &   \fr{ 2\sqrt{3}\ a_J}{3 N_c} &   - \fr{  \sqrt{3}}{2 N_c}   \\ [0.5ex]
\hline
\Delta_{J}  & 0 &                      0  &                           0  \\
\Sigma_{J}  & 1 &  \fr{ \sqrt{3}\ b_J}{2 N_c} &   - \fr{ 5 \sqrt{3}}{8 N_c}  \\
\Xi_{J}     & 2 &  \fr{ \sqrt{3}\ b_J}{  N_c} &   - \fr{ 5 \sqrt{3}}{4 N_c}  \\
\Omega_{J}  & 3 &  \fr{3\sqrt{3}\ b_J}{2 N_c} &   - \fr{15 \sqrt{3}}{8 N_c}  \\ [0.5ex]
\hline
\Sigma_{7/2}^8 - \Sigma^{10}_{7/2}    &  0 &  -\fr{  \sqrt{35}}{2 \sqrt{3} N_c}   &   0   \\
\Sigma_{9/2}^8 - \Sigma^{10}_{9/2}    &  0 &  -\fr{  \sqrt{11}}{  \sqrt{3} N_c}   &   0   \\
\Xi_{7/2}^8    - \Xi^{10}_{7/2}       &  0 &  -\fr{  \sqrt{35}}{2 \sqrt{3} N_c}   &   0   \\
\Xi_{9/2}^8    - \Xi^{10}_{9/2}       &  0 &  -\fr{  \sqrt{11}}{  \sqrt{3} N_c}   &   0   \\[0.5ex]
\hline
\hline
\end{array}
\]
\caption{Matrix elements of SU(3) breaking operators, 
with  $a_J = 5/2,-2$ for $J=7/2, 9/2$ respectively and
$b_J = 5/2, 4/3, -1/6, -2$ for $J=5/2, 7/2, 9/2, 11/2$, respectively.}
\label{octets}
\end{table}


\begin{table}
\[
\begin{array}{crcl}
\hline
\hline
(1) & 9(\Delta_{7/2} - \Delta_{5/2})     & = & 7(N_{9/2} - N_{7/2})  \\
(2) & 9 (\Delta_{9/2} - \Delta_{5/2}) & = & 16 (N_{9/2} - N_{7/2})  \\
(3) & 9(\Delta_{11/2} - \Delta_{9/2}) & = & 11 (N_{9/2} - N_{7/2})  \\
\hline
(4) & 8 (\Lambda_{7/2} - N_{7/2}) +14 (N_{9/2}-\Lambda_{9/2})
&=&   3 (\Lambda_{9/2}-\Sigma_{9/2}) +6 (\Delta_{11/2}-\Sigma_{11/2})  \\
(5) & \Lambda_{9/2} - \Lambda_{7/2} + 3(\Sigma_{9/2} - \Sigma_{7/2}) & = &  4 (N_{9/2} - N_{7/2}) \\
(6) & \Lambda_{9/2} - \Lambda_{7/2} + \Sigma_{9/2} - \Sigma_{7/2} & = &  2 (\Sigma'_{9/2} - \Sigma'_{7/2})  \\
(7) & 11 \; \Sigma'_{7/2} + 9 \; \Sigma_{11/2} &=& 20 \; \Sigma'_{9/2}   \\
(8) & 20 \; \Sigma_{5/2} + 7 \; \Sigma_{11/2} &=& 27 \; \Sigma'_{7/2} \\
\hline
{\rm (GMO)  }    &   2 (N + \Xi) &=& 3 \; \Lambda + \Sigma  \\
{\rm (EQS)  }       &  \Sigma - \Delta &=& \Xi - \Sigma = \Omega - \Xi  \\
\hline
\hline
\end{array}
\]
\caption{The 18 independent mass relations including
the GMO relations for the octets and the EQS relations for the decuplets.}
\label{masrel}
\end{table}

\pagebreak


\begin{table}
\begin{tabular}{crrrrccl}\hline \hline
                    &      \multicolumn{6}{c}{1/$N_c$ expansion results}        &                     \\ 
\cline{1-6}		    
                    &      \multicolumn{4}{c}{Partial contribution (MeV)} & \hspace{.4cm} Total (MeV)  \hspace{.4cm}  & \hspace{.2cm}  Empirical \hspace{.2cm} &  Name, status \\
\cline{2-5}
                    &   \hspace{.3cm}   $c_1O_1$  & \hspace{.3cm}  $c_2O_2$ & \hspace{.3cm}$c_3O_3$ &\hspace{.3cm}  $b_1\bar B_1$   & \hspace{.4cm}   &  (MeV)   &    \\
\hline
$N_{7/2}$        & 2209 & -3 &  34 &   0  &  \hspace{.4cm} $ 2240\pm97 $ \hspace{.4cm} &\hspace*{.2cm}  \hspace{.2cm} &  \\
$\Lambda_{7/2}$  &      &     &    & 110  &  $2350\pm118 $  & & \\
$\Sigma_{7/2}$   &      &     &    & 110  &  $2350\pm118 $  &               & \\
$\Xi_{7/2}$      &      &     &    & 220  &  $2460\pm 166$  &               &  \\
\hline
$N_{9/2}  $      & 2209 & 2   & 34 &   0  &   $2245\pm95 $  & $ 2245\pm65 $ & N(2220)**** \\
$\Lambda_{9/2}$  &  &     &    & 110  &  $ 2355\pm116 $  & $ 2355\pm15 $ &  $\Lambda$(2350)***\\
$\Sigma_{9/2}$   &      &     &    & 110  &  $ 2355\pm116 $  &               &                    \\
$\Xi_{9/2}$      &      &     &    & 220  &  $2465\pm164$  &               &                    \\
\hline
$\Delta_{5/2}$   & 2209 & -9  &168 &   0  &  $ 2368\pm175$  & &  \\
$\Sigma^{}_{5/2}$&      &     &    & 110  & $2478\pm187$  & &   \\
$\Xi^{}_{5/2}$   &      &     &    & 220  &  $2588\pm220$  &               &                   \\
$\Omega_{5/2}$   &      &     &    & 330  & $2698\pm266$  &               &                    \\
\hline
$\Delta_{7/2}$   &2209  &-5   &168 &  0   &  $2372\pm153$  & $2387\pm88$ &  $\Delta$(2390)* \\
$\Sigma'_{7/2}$  &      &     &    & 110  & $2482\pm167$  &               &                  \\
$\Xi'_{7/2}$     &      &     &    & 220  & $2592\pm203$  &               &                    \\
$\Omega_{7/2}$   &      &     &    & 330  &  $2702\pm252$  &               &                    \\
\hline
$\Delta_{9/2}$   &2209  & 1   &168 &  0   &   $2378\pm144 $  & $2318\pm132  $ &  $\Delta$(2300)**\\
$\Sigma'_{9/2}$  &      &     &    & 110  &   $2488\pm159$  &               &                   \\
$\Xi'_{9/2}$     &      &     &    & 220  &  $2598\pm197$  &               &                    \\
$\Omega_{9/2}$   &      &     &    & 330  &  $2708\pm247$  &               &                    \\
\hline
$\Delta_{11/2}$  &2209  &7    &168 &  0   &  $2385\pm164$  & $ 2400\pm100$ &   $\Delta$(2420)**** \\
$\Sigma^{}_{11/2}$ &    &     &    & 110  & $2495\pm177$  &               &                     \\
$\Xi^{}_{11/2}$  &      &     &    & 220  &  $2605\pm212$  &               &                     \\
$\Omega_{11/2}$  &      &     &    & 330  &  $2715\pm260$  &               &                     \\
\hline
\hline
\end{tabular}
\caption{The partial contribution and the total mass (MeV)
predicted by the $1/N_c$ expansion as compared with the 
empirically known masses.}
\label{multiplet}
\end{table}

\begin{table}[tbp]
{\renewcommand{\arraystretch}{1.5}
\begin{tabular}{lrr}
\hline \hline
Baryon         &   \hspace{0cm}  $\phi^{\lambda}$     & \hspace{0cm} $\phi^{\rho}$ \\
\hline
$p$              & \hspace{0cm}$-\frac{1}{\sqrt{6}}\left( udu+duu-2uud\right)$ & \hspace{0cm}$\frac{1}{\sqrt{2}}\left( udu-duu\right)$\\
$n$          & $\frac{1}{\sqrt{6}}\left( udd+dud-2ddu\right)$ & $\frac{1}{\sqrt{2}}\left( udd-dud\right)$\\
$\Sigma^+$ & $\frac{1}{\sqrt{6}}\left( usu+suu-2uus\right)$ & $-\frac{1}{\sqrt{2}}\left( usu-suu\right)$\\
$\Sigma^0$ & {\renewcommand{\arraystretch}{0.4}\begin{tabular}[t]{c}
$-\frac{1}{\sqrt{12}}\left( 2uds+2dus\right.$ \\ $\left. -sdu-sud-usd-dsu\right)$\end{tabular}} & \hspace{0.5cm}$-\frac{1}{2}\left( usd+dsu-sdu-sud\right)$\\
$\Sigma^-$  & $\frac{1}{\sqrt{6}}\left( dsd+sdd-2dds \right)$ & $-\frac{1}{\sqrt{2}}\left( dsd-sdd\right)$ \\
$\Lambda^0$ & $\frac{1}{2}\left( sud-sdu+usd-dsu \right)$ & {\renewcommand{\arraystretch}{0.4}\begin{tabular}[t]{c} $\frac{1}{\sqrt{12}}\left( 2uds-2dus \right.$ \\ $\left. +sdu-sud+usd-dsu\right)$\end{tabular}}\\ 
$\Xi^0$  & $-\frac{1}{\sqrt{6}}\left( uss+sus-2ssu \right)$  & $-\frac{1}{\sqrt{2}}\left( uss-sus\right)$ \\
$\Xi^-$  & $-\frac{1}{\sqrt{6}}\left( dss+sds-2ssd \right)$ & $-\frac{1}{\sqrt{2}}\left( dss-sds\right)$\\
\hline \hline
\end{tabular}}
\caption{Mixed symmetry flavor states of three quarks for the baryon octet.}
\label{msymflstates}
\end{table}

\begin{table}[tbp]
{\renewcommand{\arraystretch}{1.5}
\begin{tabular}{lr}
\hline \hline
Baryon  &   $\phi^S$ \\
\hline
$\Delta^{++}$ & $uuu$ \\
$\Delta^{+}$ & $\frac{1}{\sqrt{3}}\left( uud+udu+duu \right)$ \\
$\Delta^0$   & $\frac{1}{\sqrt{3}}\left( udd+dud+ddu \right)$ \\
$\Delta^-$   & $ddd$ \\
$\Sigma^+$ & $\frac{1}{\sqrt{3}}\left( uus+usu+suu \right)$ \\
$\Sigma^0$ & $\frac{1}{\sqrt{6}}\left( uds+dus+usd+sud+sdu+dsu\right)$\\
$\Sigma^-$ & $\frac{1}{\sqrt{3}}\left( sdd+dsd+dds\right)$\\
$\Xi^0$ & $\frac{1}{\sqrt{3}}\left( uss+sus+ssu\right)$\\
$\Xi^-$ & $\frac{1}{\sqrt{3}}\left( dss+sds+ssd\right)$\\
$\Omega^-$& $sss$\\
\hline \hline
\end{tabular}}
\caption{Symmetric flavor states of three quarks for the baryon decuplet.}
\label{symflstates}
\end{table}

\end{document}